\documentclass[sigconf]{acmart}
\usepackage{subfigure}
\usepackage{diagbox}
\usepackage{multirow}
\usepackage[normalem]{ulem}
\usepackage{algorithm}
\usepackage{algorithm}
\usepackage{algorithmicx}
\usepackage{algpseudocode}
\useunder{\uline}{\ul}{}
\usepackage{float}
\usepackage{xcolor}

\AtBeginDocument{%
  \providecommand\BibTeX{{%
    \normalfont B\kern-0.5em{\scshape i\kern-0.25em b}\kern-0.8em\TeX}}}
    
\title{Contextual Distillation Model for Diversified Recommendation}

\author{Fan Li}
\authornote{Equal contributions.} 
\affiliation{%
  \institution{University of Science and Technology of China}
  \city{Hefei}
  \country{China}
}
\email{lf321@mail.ustc.edu.cn}

\author{Xu Si}
\authornotemark[1] 
\affiliation{%
  \institution{Tsinghua University}
  \city{Beijing}
  \country{China}
}
\email{six21@mails.tsinghua.edu.cn}

\author{Shisong Tang}
\affiliation{%
  \institution{Kuaishou Inc.}
  \institution{Tsinghua University}
  \city{Beijing}
  \country{China}
}
\email{tangshisong@kuaishou.com}

\author{Dingmin	Wang}
 \affiliation{%
  \institution{University of Oxford}
  \city{Oxford}
  \country{United Kingdom}
}
\email{dingmin.wang@cs.ox.ac.uk}

\author{Kunyan Han}
\affiliation{%
  \institution{Kuaishou Inc.}
  \city{Beijing}
  \country{China}
}
\email{hankunyan@kuaishou.com}

\author{Bing Han}
\affiliation{%
  \institution{Kuaishou Inc.}
  \city{Beijing}
  \country{China}
}
\email{hanbing@kuaishou.com}

\author{Guorui Zhou}
\affiliation{%
  \institution{Kuaishou Inc.}
  \city{Beijing}
  \country{China}
}
\email{zhouguorui@kuaishou.com}

\author{Yang Song}
\affiliation{%
  \institution{Kuaishou Inc.}
  \city{Beijing}
  \country{China}
}
\email{yangsong@kuaishou.com}

\author{Hechang Chen}
\authornote{Corresponding author.}
\affiliation{%
  \institution{Jilin University}
  \city{Changchun}
  \country{China}
}
\email{chenhc@jlu.edu.cn}

\copyrightyear{2024} 
\acmYear{2024}
\setcopyright{acmlicensed}
\acmConference[KDD '24]{Proceedings of the 30th ACM SIGKDD Conference on Knowledge Discovery and Data Mining}{August 25--29, 2024}{Barcelona, Spain}
\acmBooktitle{Proceedings of the 30th ACM SIGKDD Conference on Knowledge Discovery and Data Mining (KDD '24), August 25--29, 2024, Barcelona, Spain} \acmDOI{10.1145/3637528.3671514} 
\acmISBN{979-8-4007-0490-1/24/08}
 
\renewcommand{\shortauthors}{Fan Li et al.}

\begin{document}

\begin{abstract}

The diversity of recommendation is equally crucial as accuracy in improving user experience.
Existing studies, e.g., Determinantal Point Process (DPP) and Maximal Marginal Relevance (MMR), employ a greedy paradigm to iteratively select items that optimize both accuracy and diversity.
However, prior methods typically exhibit quadratic complexity,  limiting their applications to the \textit{re-ranking} stage and are not applicable to other recommendation stages with a larger pool of candidate items, such as the \textit{pre-ranking} and \textit{ranking} stages.
In this paper, we propose \textbf{C}ontextual \textbf{D}istillation \textbf{M}odel (CDM), an efficient recommendation model that addresses diversification, suitable for the deployment in all stages of industrial recommendation pipelines.  
Specifically, CDM utilizes the candidate items in the same user request as context to enhance the diversification of the results. 
We propose a contrastive context encoder that employs attention mechanisms to model both positive and negative contexts.
For the training of CDM, we compare each target item with its context embedding and utilize the knowledge distillation framework to learn the win probability of each target item under the MMR algorithm, where the teacher is derived from MMR outputs. 
During inference, ranking is performed through a linear combination of the recommendation and student model scores, ensuring both diversity and efficiency.
We perform offline evaluations on two industrial datasets and conduct online \textit{A}/\textit{B} test of CDM on the short-video platform \textit{KuaiShou}. 
The considerable enhancements observed in both recommendation quality and diversity, as shown by metrics, provide strong superiority for the effectiveness of CDM.


\end{abstract}

\ccsdesc[500]{Information systems~Recommender systems}
\keywords{Recommender System, Knowledge Distillation, Diversified Recommendation} 
\maketitle

\section{Introduction}
With the prevalence of Web 2.0 and mobile devices, an increasing number of users are joining online feed stream platforms such as TikTok\footnote{\href{https://www.tiktok.com/}{https://www.tiktok.com/}}, Douyin\footnote{\href{https://www.douyin.com/}{https://www.douyin.com/}}, and KuaiShou\footnote{\href{https://www.kuaishou.com/new-reco}{https://www.kuaishou.com/new-reco}} for content sharing and consumption \cite{shisong2022,shisong2023,fu1,fu2,sun1}.
To alleviate the issue of content homogenization and enhance user engagement, it is important to consider the diversity in recommendation systems. 
The diversity quantifies the dissimilarity among recommended items based on certain pre-selected attributes (e.g., item category),  which is as crucial as accuracy. Enhanced diverity can significantly contribute to the overall user experience by encouraging exploration of new content and discovery of new interests \cite{wang2017deep_dcn, zhang2023disentangled, wu2019_div_survey, wilhelm2018practical,kapoor2015like}.

\begin{figure}[t]
\includegraphics[scale=0.45]{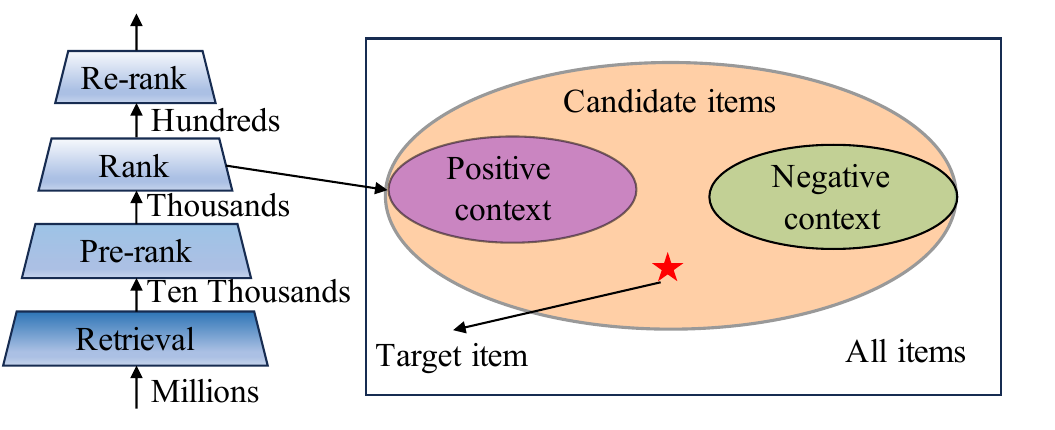}
\caption{An industry recommendation pipeline and different item sets for one user request. Candidate items refer to: the set of items that are to be scored one user request during the \textit{ranking} stage.}
\label{fig:pipeline}
\end{figure}

Figure \ref{fig:pipeline} illustrates the stages of an industrial recommendation pipeline. 
Existing research on diversified recommendation \cite{carbonell1998use_mmr,chen2018fast_dpp,huang2021sliding_ssd,xu2023multi}, such as the Determinantal Point Process (DPP) \cite{chen2018fast_dpp} and Maximal Marginal Relevance (MMR) \cite{carbonell1998use_mmr}, has predominantly focused on enhancing diversity in the \textit{re-ranking} stage. 
These greedy-based approaches have successfully augmented the diversity of \textit{re-ranking} recommendation model. 
However, the necessity for diversity extends beyond the \textit{re-ranking} stage, encompassing earlier stages such as \textit{pre-ranking} and \textit{ranking}.
The absence of diversity in these preliminary stages will render subsequent efforts to enhance diversity during \textit{re-ranking} ineffective against the backdrop of a highly homogenized candidate item pool, leading to a predicament metaphorically described as "You can't make bricks without straw." 
Therefore, ensuring the diversity of the entire pipeline stages is crucial in industrial recommendation pipeline.

However, the existing diversity algorithms designed for the \textit{re-ranking} stage are characterized by quadratic computational complexity, making them unsuitable for stages involving extensive candidate item pools, such as \textit{pre-ranking} and \textit{ranking}, due to their prohibitively high computational demands. 
Consequently, there is a pressing need to develop an algorithm that is not only efficient in handling large sets of candidates but also capable of effectively enhancing diversity during the \textit{pre-ranking} and \textit{ranking} stages, thereby significantly elevating the overall efficacy of recommendation pipelines.

In this paper, we introduce Contextual Distillation Model (CDM), an efficient recommendation model designed to enhance diversification across all stages of the industrial recommendation pipeline.
CDM innovatively addresses the need for both efficiency and diversity by training a surrogate model of the quadratic time-complexity MMR algorithm. 
By employing the knowledge distillation framework \cite{hinton2015distilling}, the surrogate model is trained to estimate the probability of an item being selected by the MMR algorithm (i.e. the Top-K items). 
This strategy enables the deployment of an end-to-end student model that enhances diversity in the \textit{pre-ranking} and \textit{ranking} stages in an efficient manner.

\begin{figure}[t]
\includegraphics[scale=0.43]{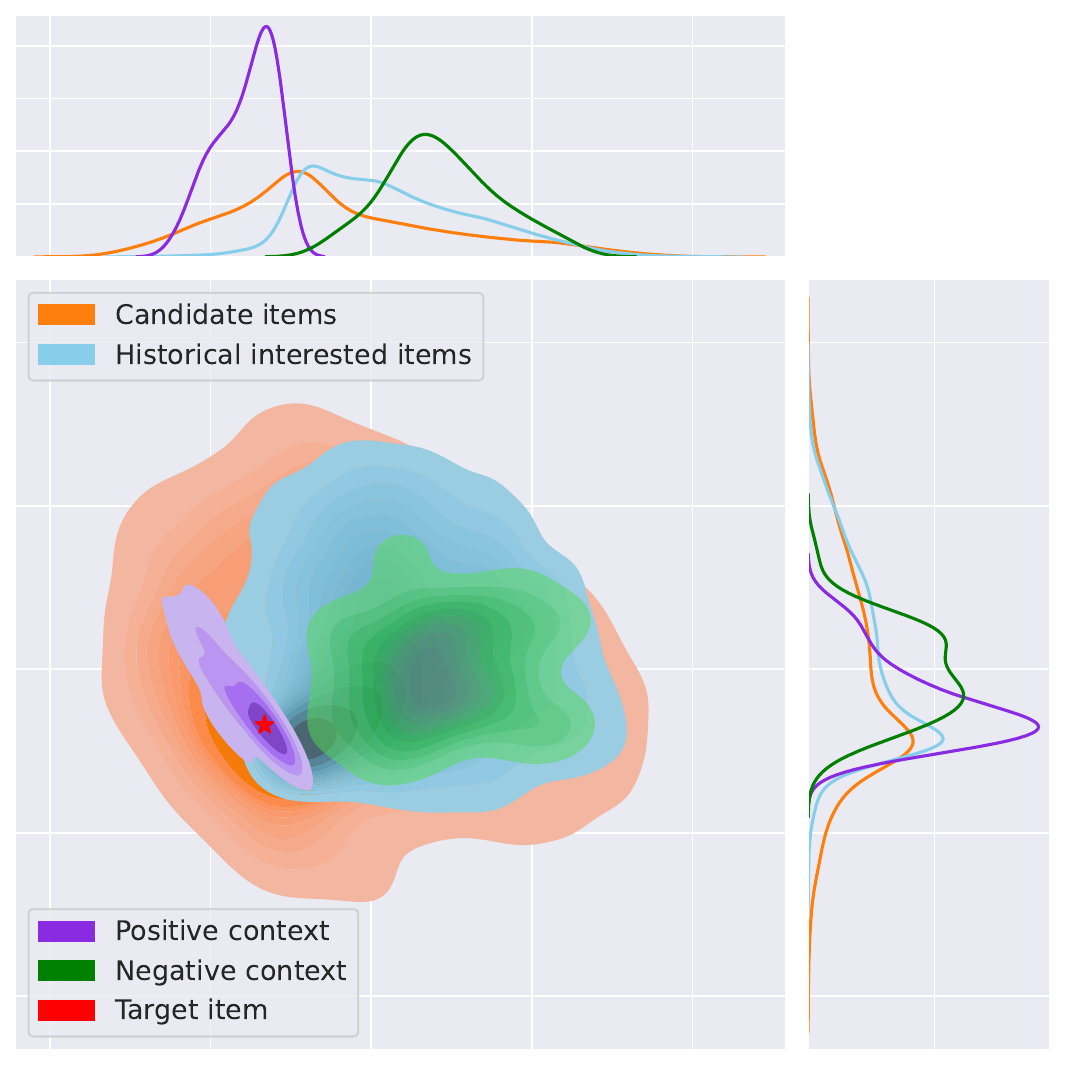}
\caption{The probability density distribution of a target item, where randomly selected from a candidate set in a single user request, corresponds to different sets of item embeddings. The positive context comprises the 100 candidate items most similar to target item, and the negative context, the least similar.}
\label{fig:distribution}
\end{figure}

The next problem is how to architecture the student model.
We know that diversity serves as a metric for evaluating sets.
However, models in \textit{pre-ranking} and \textit{ranking} stages traditionally score each item independently (point-wise).
Therefore, to introduce set information into the scoring process, CDM leverages the candidate items within the same user request as context to enhance the diversity of recommendations.
Specifically, we propose a Contrastive Context Encoder (CCE) to model the context for each target item. 
Given the large number of candidates in \textit{pre-ranking} and \textit{ranking} stages, it's impractical to incorporate all context into calculations with the target item, necessitating a sampling work on the context.
We posit that for any given target item, there exist two distinct types of contexts within the candidate pool: highly similar items (positive context) and markedly dissimilar items (negative context).
Figure \ref{fig:distribution} shows the probability density distribution of a target item, where randomly selected from a candidate set in a single user request, corresponds to different sets of item embeddings.
It is essential to sample these two contexts, executed by employing a Gumbel-Top-$k$ strategy \cite{xie2019reparameterizable} based on the target attention scores \cite{vaswani2017attention} between the target item and the context.
After that, we employ the contrastive attention mechanism \cite{song2018mask, duan2019contrastive} to effectively model the distinguishing representations between the two opponent contexts, enhancing the model's ability to capture diverse patterns.
This mechanism, initially introduced in the field of computer vision \cite{song2018mask}, has been primarily utilized for person re-identification by contrastively attending to person and background regions.

Ultimately, by comparing each target item with its context embedding, we can discern whether an item surpasses the aggregate performance, thereby estimating its probability of being selected in the MMR algorithm \cite{carbonell1998use_mmr}. 
During the inference phase, ranking is performed through a linear combination of scores from both the recommendation and student models, ensuring diversity and efficiency. 
This approach not only theoretically pioneers a novel solution but also empirically validates the efficacy of CDM in elevating both the quality and diversity of recommendations through offline evaluations on two industrial datasets and an online A/B test on the short-video platform KuaiShou.

Our contributions can be summarized as follows:
\begin{itemize}
    \item We propose a Context Distillation Model for diversified recommendations, which extracts context embeddings from the candidates of pipeline for each target item, and learns end-to-end diversity scores in a distillation way.
    \item We propose a Contrastive Context Encoder to learn context embeddings from the positive and negative contextual information inherent in each target item's candidate items.
    \item Both offline experiments on two industrial datasets and online \textit{A}/\textit{B} test on the feed pages of \textit{KuaiShou} App demonstrate the effectiveness of our framework in balancing recommendation accuracy and diversity.
\end{itemize}



\section{Preliminaries}\label{sec:2.1}

\subsection{\textbf{Traditional Recommendation}}
Let $\mathcal{U} = \{u_1,...,u_M\}$ and $\mathcal{I} = \{i_1,...,i_{N}\}$ denote the set of users and  \textbf{candidate items within a single user request}, respectively, where $M$ is the number of users, $N$ is the number of candidate items. The user-item historical interactions are represented by $\mathcal{D} = \{(u,i,y)|u \in \mathcal{U},i \in \mathcal{I} \}$, where $y \in \{0,1\}$ denotes the binary label (e.g., finish playing or like). The target of the traditional recommendation training is to learn the scoring function $f(u,i|{\Theta})$ from $\mathcal{D}$, which is capable of predicting the preference of user $u$ on item $i$, where $\Theta$ is the parameters of $f$.

\subsection{\textbf{Diversified Recommendation}}

Diversified recommendation aims to provide users with a set of accurate and diverse items. It can be formally defined as a subset selection problem. For each user $u \in \mathcal{U}$, the objective is to determine a ranked list $\mathcal{R} = \{i_1, i_2, \ldots, i_K\}$ of $K$ items from the candidate set $\mathcal{I}$ that maximizes a linear combination of accuracy and diversification, which follows the classical work Maximal Marginal Relevance (MMR):

\begin{equation}
\mathcal{R} = \mathop{\mathrm{argmax}} \limits_{\mathcal{R}^{'} \subset \mathcal{I}} {\mathrm{Acc}(u,\mathcal{R}^{'}) + \lambda*\mathrm{Div}(u,\mathcal{R}^{'})}.
\end{equation}

Here, $\mathrm{Acc}(\cdot)$ represents accuracy, ensuring that the recommended items align with the user's preferences and have a high likelihood of interaction. On the other hand, $\mathrm{Div}(\cdot)$ quantifies diversity, ensuring that the recommended items are dissimilar, offering a range of choices to the user. The parameter $\lambda$ controls the trade-off between accuracy and diversity.

\subsection{\textbf{Approximation Calculation}} 
Our current challenge revolves around a discrete optimization problem. Specifically, it is tasked to select a subset $\mathcal{R}$ of size $K$ from candidate items $\mathcal{I}$ to optimize MMR. However, existing research has demonstrated that maximizing a submodular function with cardinality constraints constitutes an NP-hard problem \cite{toth2000optimization}. Some works have designed $(1-1/e)$-approximation algorithms for this problem based on greedy techniques \cite{carbonell1998use_mmr, minoux2005accelerated}. That is, the construction of $\mathcal{R}$ follows an iterative process according to marginal gain:
\begin{equation}\label{i_k}
i_h = \mathop{\mathrm{argmax}} \limits_{i \in \mathcal{I} \textbackslash \mathcal{R}_{1:h-1}} {\mathrm{Acc}(u,i) + \lambda*\mathrm{Div}(\mathcal{R}_{1:h-1} \cup {i} | u)},
\end{equation}
where $h$ represents the current selection step, and our choices are influenced by the outcomes of the preceding  $h-1$ steps. 

However, the greedy selection's quadratic complexity, at $O(K^2N)$, limits its utility primarily to the \textit{re-ranking} stage. 
Such high computational demands render it impractical for other recommendation stages, notably the \textit{pre-ranking} and \textit{ranking} stages, where a larger pool of candidate items necessitates more scalable solutions.
Consequently, our objective is to achieve the benefits of the greedy approach's optimal results while significantly reducing the computational overhead in these more extensive stages.

\begin{algorithm}[h]
  \caption{Maximum Marginal Relevance Algorithm}
  \begin{algorithmic}[1]
       \State $\mathcal{R} = \{\mathop{\mathrm{argmax}}_{i\in \mathcal{I}} \mathrm{Acc}(u,i)\}$
       \While {$|\mathcal{R}| < K$}
         \For {$i \in \mathcal{I} \setminus \mathcal{R}$}
           \State $\mathcal{R} = \mathcal{R} \cup \{\mathop{\mathrm{argmax}} \left(\mathrm{Acc}(u,i) + \lambda*\mathrm{Div}(\mathcal{R} \cup \{i\} | u)\right)\}$
         \EndFor
       \EndWhile
  \end{algorithmic}
  \label{alg}
\end{algorithm}

\section{Method}

\subsection{Interest-Aware MMR}

The fundamental technique for diversity ranking is known as Maximum Marginal Relevance (MMR) \cite{carbonell1998use_mmr}, and considerable researches \cite{lin2022feature,wu2019_div_survey} have been dedicated to enhancing and extending algorithms grounded in MMR principles. 
In this section, we present an enhancement of MMR that incorporates user interests, thereby expanding the scope of diversity considerations. 
For the accuracy score $\text{ACC}(u,i)$, we directly employ the score function $f(u,i)$. 
To compute the diversity score, we incorporate the user interest in the measurement of item similarity. 
This incorporation links diversity measurement to the user's personal perception of diversity stemming from distinct interests. 
Specifically,
\begin{equation}
    \mathrm{Div}(\mathcal{R}_{1:h-1} \cup {i} | u) = 1-\max_{i^{'}\in \mathcal{R}_{1:h-1}} \sigma(\mathbf{e}_{i,u}\cdot \mathbf{e}_{i{'},u}),
\end{equation}
where $\mathbf{e}_{i,u} = \mathbf{e}_{i} \odot \mathbf{u}$ refers to the Hadamard product between the item embedding $\mathbf{e}_{i}$ and the user interest vector $\mathbf{u}$. In this diversity function, we subtract the maximal similarity between selected items $\mathcal{R}$ and the target item $i$ as the diversity score.

Let's begin by addressing the selection of the initial item, referred to as $i_1$. In this simple case, where no items have been selected yet (i.e., $\mathcal{R} = \emptyset$), it simplifies to choosing the item with the highest single-item accuracy score, as expressed by:
\begin{equation}
    i_1 = \mathop{\mathrm{argmax}}_{i\in \mathcal{I}} f(u,i).
\end{equation}
Subsequently, the selection of the remaining $K-1$ items is carried out iteratively following Eq. \ref{i_k}. 
The top-K items that can be selected as positive samples, with the rest as negative samples, guide the training of the surrogate model using this as the label.


\subsection{Contrastive Context Learning}
Previously, we discussed the employment of MMR's outputs as supervised signals for training a surrogate model, denoted as $g(\cdot)$, to learn diversity. 
Regarding the input to $g(\cdot)$, we employ the candidate items generated by the \textit{pre-ranking} stage, see Fig \ref{fig:pipeline}. 
In this section, we focus on how to learn the context embeddings from the candidates of target item for serving as the inputs to $g(\cdot)$ in the \textit{ranking} stage.

\begin{figure}[t]
\includegraphics[scale=0.64]{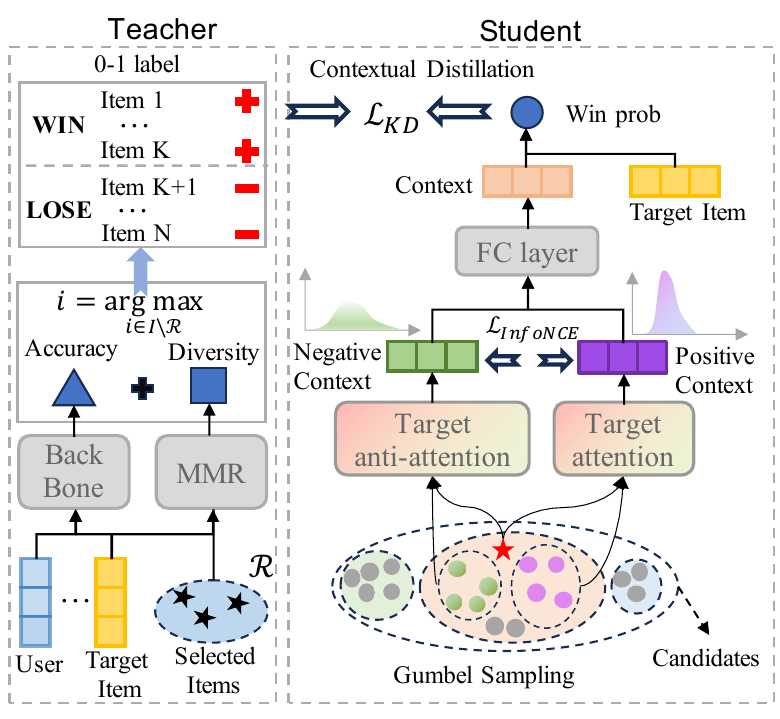}
\caption{The architecture of our proposed CDM. 
The teacher model is the MMR algorithm, and the student model predicts each item's winning probability with context.}
\label{fig:model}
\end{figure}

\subsubsection{\textbf{Differentiable Sampling with Gumbel-Top-$k$ Reparameterization.}}
During the \textit{ranking} stage of industrial recommendation, the size of candidate items $N$ for a single user request is considerable, as shown in Fig \ref{fig:pipeline}. 
Incorporating all candidate items into graph computations brings additional overhead and increases undesirable noise that affects context learning. 
Opting for sampling emerges as an efficacious solution.
However, simple random sampling falls short in resolving the aforementioned issues. 
Upon delving into the item distribution within each user request, we discover a significant pattern: each target item invariably has both relevant and competing counterparts within the candidate items. 
Interestingly, the distribution of relevant items tends to be densely clustered, while that of competing ones exhibit a looser distribution, as shown in Fig \ref{fig:distribution}. 
In essence, this observation motivates us to sample from these two distributions separately, thereby modeling both positive and negative context information for each target item.

Given a candidate item set $\mathcal{I}^t = \{i_1, \ldots, i_m, \ldots, i_{N}\}$ and a target item $i^t$, we compute the attention score $w_m = s(i^t,i_m)$ between target item $i^t$ and each candidate item $i_m$. However, the sampling of $\mathbf{w}$ is non-differentiable and prevents the model from being trained well via back-propagation. To solve this issue, we adopt a differentiable ordered subset sampling without \textit{replacement} by generalizing the Gumbel-Softmax reparametrization \cite{jang2016categorical} to the Gumbel-Top-$k$ trick \cite{xie2019reparameterizable}. We define the probability of $w_m$ as $p(i_m|i^t) = p_m$. The perturbed weights are then obtained by adding Gumbel noise $g$ to the log probabilities $p$:
\begin{equation}
    \widetilde{w}_m = \log p_m + g_m,
\end{equation}
\begin{equation}
    g_m = -\log(-\log(u_m)) \sim \text{Gumbel}(0,1),
\end{equation}
where $u_m \sim \text{Uniform(0,1)}$. Now, we need to consider how to turn the perturbed weights $\widetilde{\mathbf{w}} = [\widetilde{w}_1,... ,\widetilde{w}_{N}]$ into one-hot vectors $\textbf{a} = [\text{a}_1,...,\text{a}_{N}]$, s.t. $\sum_{m^{'}}  \text{a}_{m^{'}} = k$. 

Following the work \cite{plotz2018neural}, we propose to use a recent top-$k$ relaxation based on successive applications of the softmax function. Specifically, define for all candidate item indices $m = 1,...,N$ and iteration steps $j = 1,...,k$:
\begin{equation}\label{eq:hard}
    \alpha_m^{j+1} = \alpha_m^j + \log(1-\text{a}^j_m), \quad\quad \alpha_m^1 = \widetilde{w}_m,
\end{equation}
where $\alpha_m^j$ is a sample distribution at iteration step $j$.

When considering $\text{a}^j_m$, \cite{plotz2018neural} proposes a replacement using its expectation in a relaxed manner, i.e.:
\begin{equation}
    \mathrm{P}(\text{a}_m^{j}=1) = \frac{\exp{(\alpha_m^{j}/\tau)}}{\sum_{m^{'}} \exp{(\alpha_{m^{'}}^{j}/\tau)}},
\end{equation}
where $\tau > 0$ is a given temperature hyper-parameter, such that the new update is:
\begin{equation}\label{eq:soft}
    \alpha_m^{j+1} = \alpha_m^j + \log(1-\mathrm{P}(\text{a}_m^{j}=1)). 
\end{equation}
In this way, each step $j$ produces a relaxed one-hot vector $\textbf{a}^j = [\mathrm{P}(\text{a}_1^{j}=1),...,\mathrm{P}(\text{a}_{N}^{j}=1)]$, and the output $\textbf{a} = \sum_{j=1}^k \textbf{a}^j$ is the relaxed $k$-hot vector, i.e.:
\begin{equation}
    \textbf{a} = \left[\sum_{j=1}^k \mathrm{P}(\text{a}_1^{j}=1),...,\sum_{j=1}^k \mathrm{P}(\text{a}_{N}^{j}=1)\right].
\end{equation}
Previous work \cite{plotz2018neural} has demonstrated that as the temperature hyper-parameter $\tau \rightarrow 0$, the expectation $\textbf{a}^1=[\mathrm{P}(\text{a}_1^{1}=1),...,\mathrm{P}(\text{a}_{N}^{1}=1)]$ of the initial sampled vector $\widetilde{\textbf{w}}$ approximates a one-hot encoding. Consequently, the logit update in Eq.(\ref{eq:soft}) will also converge to the hard update from Eq.(\ref{eq:hard}). By induction, this convergence trend extends to others, resulting in a one-hot encoding. Significantly, it aligns with the Gumbel-Softmax reparametrization \cite{jang2016categorical} when $k=1$. In the end, the output relaxed $k$-hot vector $\textbf{a}$ transforms into a hard encoding that satisfies the desired property of $\sum_{m^{'}}  \text{a}_{m^{'}} = k$ as $\tau \rightarrow 0$. To counterbalance the influence of model initialization and broaden parameter exploration, we initiate model training with a comparatively larger value of $\tau$ for expectation computation. Subsequently, we implement a gradual annealing process \cite{xu2015low} that progressively reduces $\tau$ to a small constant value during training.

Now, we can employ the Gumbel-Top-$k$ trick to sample $k$ relevant candidate items $\mathcal{I}_t^{pos} = \{i_1^{pos},i_2^{pos},...,i_k^{pos}\}$ as positive contexts, and $k$ competing candidate items $\mathcal{I}_t^{neg} = \{i_1^{neg},i_2^{neg},...,i_k^{neg}\}$ as negative contexts for each target item $i^t$ from its corresponding candidate item set $\mathcal{I}$.

\subsubsection{\textbf{Contrastive Context Encoder}}
Prior to the sampling phase, the attention score 
$\mathbf{w}$ is computed between the target item and its candidate items as follows \cite{vaswani2017attention, yu2020tagnn}:
\begin{equation}
    \mathbf{w} = \frac{\mathbf{q}^{T}\mathbf{K}}{\sqrt{d}} \in \mathbb{R}^{N},
\end{equation}
where $\mathbf{q} = i^t\cdot \mathbf{W}_1$ represents the embedding of the target item and $\mathbf{K} = [i_1,...,i_{N}]*\mathbf{W}_2$ corresponds to the embeddings of candidate items. $\mathbf{W}_1$ and $\mathbf{W}_2$ are two learnable parameters.

Following the sampling procedure, we acquire two distinct contexts: a positive context $\mathcal{I}_t^{pos}$ and a negative one $\mathcal{I}_t^{neg}$. The subsequent task involves sequences modeling. In this regard, we utilize the \textit{contrastive attention} \cite{song2018mask, duan2019contrastive} to directly model two sequences and generate the corresponding context embeddings. The computation of positive context embedding is used \textit{target attention} \cite{vaswani2017attention, yu2020tagnn}:
\begin{equation}
    \mathbf{C}^{pos} = \text{softmax}(\mathbf{\widetilde{w}}^{pos})\cdot \mathbf{V}^{pos},
\end{equation}
where $\mathbf{\widetilde{w}}^{pos}$ signifies the perturbed attention weights of the positive candidate items, generated during the sampling phase, and $\mathbf{V}^{pos} = [i_1^{pos},...,i_k^{pos}]*\mathbf{W}_3$ represents the candidate item's embedings. $\mathbf{W}_3$ denotes the learnable parameter. Conversely, the negative context embedding is constructed using an inverse weighting approach through \textit{target anti-attention}:
\begin{equation}
    \mathbf{C}^{neg} = \text{softmax}(-\mathbf{\widetilde{w}}^{neg})\cdot \mathbf{V}^{neg},
\end{equation}
where $\mathbf{V}^{neg} =[i_1^{neg},...,i_k^{neg}]*\mathbf{W}_3$ utilizes the same parameters as used in target-attention.

The separation of positive and negative context embeddings stems from their distinct roles in relation to the target item, aiming to precisely model the set information of the target item. Our objective is to ensure that the positive context embedding $\mathbf{C}^{pos}$ accurately captures the semantic essence associated with the target item, whereas the negative one $\mathbf{C}^{neg}$ remains devoid of any connection to the target item's attributes. In practice, we leverage the InfoNCE loss \cite{oord2018representation} function to reinforce such differentiation.
\begin{equation}
    \mathcal{L}_{InfoNCE} = -\log \frac{\exp{(\mathbf{q}^T \cdot \mathbf{C}^{pos}/t)}}{\sum_{\mathbf{C}^{'} \in\{\mathbf{C}^{pos},\mathbf{C}^{neg}\}}  \exp(\mathbf{q}^T\cdot \mathbf{C}^{'}/t)},
\end{equation}
where $t$ is a temperature parameter of InfoNCE loss.

In the end, we introduce a user interest-aware fusion strategy. Our intention is to equip the learned context embedding with the ability to discern its position within the wider user interest distribution. The final context embedding can be computed as follows:
\begin{equation}
    \mathbf{C} = \text{FFN}(\text{concat}(\mathbf{u} \odot \mathbf{C}^{pos},\mathbf{u} \odot \mathbf{C}^{neg})).
\end{equation}




\subsection{Contextual Distillation Model}

In this section, we elaborate on how the Contextual Distillation Model (Student) distills knowledge from the greedy-based Maximum Marginal Relevance (Teacher) to achieve low-cost diversified recommendations. 
We acknowledge that items selected by the MMR algorithm (i.e., the winning item set, $\mathcal{R}$), are either from niche tags or represent the finest items within popular tags. The student model contrasts the target item with its associated context to discern its alignment with the majority. Consequently, estimating the win probability for each item facilitates the guidance of the student model's learning process.
\begin{equation}
    y_{tea} = \mathbb{I}(i^t \in \mathcal{R})
\end{equation}
Regarding the student model $g(\cdot)$, it introduces the Contrastive Context Encoder, functioning as a surrogate model aimed at predicting the win probability of each target item. 
We directly compute the dot product between the embedding $\mathbf{q}$ of each target item $i^t$ and its corresponding context embedding $\mathbf{C}$ to obtain the win probability.
\begin{equation}
    y_{stu} =  g(u,i^t,\mathcal{I}) =  \sigma(\mathbf{q}^T \cdot \mathbf{C})
\end{equation}
For training $g(\cdot)$, we resort to the knowledge distillation \cite{hinton2015distilling}. The binary cross-entropy is used to minimize the discrepancy between the soft output probabilities of the student and teacher model:
\begin{equation}
    \mathcal{L}_{KD} = -y_{tea}\log y_{stu} - (1-y_{tea}) \log(1-y_{stu})
\end{equation}
Therefore, the total training loss for CDM is as follows:
\begin{equation}
    \mathcal{L}_{BCE} = -y_i\log f(u,i) - (1-y_i) \log(1-f(u,i))
\end{equation}
\begin{equation}
     \mathcal{L} =  \mathcal{L}_{BCE}+\beta_1 * \mathcal{L}_{KD} + \beta_2 * \mathcal{L}_{InfoNCE},
\end{equation}
where $\beta_1$ and $\beta_2$ are the hyper-parameters. During the inference phase, the final ranking scores are obtained by linearly fusing the accuracy and diversity scores with a hyper-parameter $\gamma$, i.e., $f(u,i)+\gamma * y_{stu}$.

\subsection{Discussion}

\subsubsection{\textbf{Time Complexity}} 
The computational efficiency of online serving is a critical concern in industrial recommender systems. 
For a single user request, where the number of candidate items during the \textit{pre-ranking} and \textit{ranking} stage is denoted as $N$, and $K$ items are to be returned. 
In the case of the MMR algorithm (list-wise), the time complexity for generating the Top-$K$ list is $O(NK^2)$. 
To elaborate, in each round, every remaining candidate item must be computed with the selected items $\mathcal{R}$. 
In contrast, the CDM approach (point-wise) provides an efficient end-to-end solution, with a time complexity equivalent to the cost of heap-sort algorithm to get the Top-$K$ list, which is $O(N \log K)$. 

It is worth noting that previous methods \cite{lin2022feature, xu2023multi} have predominantly found application in the final \textit{re-ranking} stage, typically associated with smaller $N$ and $K$ values. 
However, in the \textit{pre-ranking} and \textit{ranking} stage, where $N$ and $K$ are notably more substantial, concerns pertaining to algorithmic overhead assume heightened importance. 
CDM not only reduces the cost of inference but also enhances the original recommendation model's ability to assimilate candidate information and seamlessly integrate diversity.

\subsubsection{\textbf{Scalability}} To improve diversity within the \textit{ranking} stage, CDM employs knowledge distillation as an efficient end-to-end method for learning diversity. This approach can be applied not only to the \textit{ranking} stage but also to other stages, serving as a supplementary module for any embedding-based recommendation model. Furthermore, we have utilized MMR's score as the supervisor signal in this work, it's important to note that alternative diversity algorithms, such as the Determinantal Point Process (DPP) \cite{chen2018fast_dpp} and the Gram-Schmidt Process (GSP) \cite{huang2021sliding_ssd}, can be seamlessly substituted to tailor the diversity learning process to specific requirements.

\section{Experiment}
In this part, we first present the offline dataset and experimental configuration, as well as the reproducibility of the experiments in Section 4.1. Next, Section 4.2 and 4.3 show the evaluation results of the offline experiments as well as the detailed analysis. Finally, we show the results of A/B experiments of CDM in real scenarios in Section 4.4 to further demonstrate the effectiveness of the scheme.

\newpage
\subsection{Experimental Setup}

\subsubsection{\textbf{Dataset}} 


\begin{itemize}
    \item \textbf{JingDong} \cite{jd_context}: Constructed from search logs of a largest e-commerce platform's advertising system, this dataset encompasses features such as \textit{user id}, \textit{query id}, and item, each item containing a quintuple <\textit{item id, category id, vendor id, price id}>.
    \item \textbf{KuaiShou}\footnote{\href{https://www.kuaishou.com/new-reco}{https://www.kuaishou.com/new-reco}}: Owing to a lack of available recommendation datasets with candidate items from upstream pipelines, a new dataset was assembled from the KuaiShou app. It aggregates data from 0.5 million active users, randomly sampled, with their logs on the feed page over 5 days. The dataset incorporates \textit{user id}, \textit{tab id}, and item, each item presenting a triplet <\textit{item id, category id, duration id}>. Release of this dataset is anticipated post-review.
\end{itemize}

\begin{table}[]\caption{Statistics of two datasets.}
\begin{tabular}{c|c|c}
\toprule
Statistics         & JingDong   & Kuaishou \\ \hline
\#request          & 1,237,631  & 8M       \\
\#interaction      & 6,307,945  & 74M         \\
\#user             & 931,883    & 0.5M        \\ 
\#query            & 323,478    & -        \\
\#item             & 15,587,269 & 24M         \\
item\_show\_ratio  & 9.95\%     & 3.81\%         \\
\#avg\_candidate\_item     & 213        &  500        \\
\#tag              & 3,336      &  4k        \\ \bottomrule
\end{tabular}\label{tab:sta}
\end{table}

\subsubsection{\textbf{Metrics}.} 
For evaluation, we employ a comprehensive set of metrics to evaluate the performance from both accuracy and diversity perspectives. To measure accuracy, we utilize two widely used metrics, namely \textbf{Recall} and \textbf{Mean Reciprocal Rank (MRR)}. Regarding recommendation diversity, we incorporate two commonly used metrics: (1) \textbf{Intra List Average Distance (ILAD)}, quantifying the dissimilarity among items within the Top-$K$ list. It measures \textbf{user-level} diversity, where a larger value indicates greater diversity in the Top-$K$ recommendation list.
(2) \textbf{Category Coverage (CC)}, which is the ratio of the number of categories covered by the Top-$K$ recommendations to the total number of categories in the dataset. It focuses on \textbf{system-level} diversity, where a higher value indicates that a greater variety of categories is being recommended. The formal definitions of these metrics are as follows:

\begin{equation}
\text{ILAD@}K = 1 - \frac{1}{K(K - 1)} \sum_{x \in L} \sum_{y \in L, y \neq x} \text{Cosine}(x, y),
\end{equation}
\begin{equation}
    \text{CC@}K = \frac{|C_K|}{|C|},
\end{equation}
where $L$ denotes the Top-$K$ list, CC@$K$ represents the set of unique categories within the Top-$K$ recommendation list, and CC signifies the set of all unique categories within the dataset. Higher values of ILAD@$K$ and CC@$K$ are indicative of greater diversity in the Top-$K$ recommendations. We report the results for $K=3$ and $K=5$ on JingDong and $K=20$ and $K=50$ on KuaiShou.

\subsubsection{\textbf{Baselines}.}
\begin{itemize}
    \item \textbf{DCN} \cite{wang2017deep_dcn}: Deep Crossing Network is a state-of-the-art recommendation model serving as the backbone model of CDM.
    \item \textbf{IPW} \cite{saito2020unbiased_ipw}: During the training phase, Inverse Propensity Weighting adjusts the weights of items based on propensity scores, which are calculated from the category distribution in a user's historical interactions.
    \item \textbf{DPP} \cite{chen2018fast_dpp}: Determinantal Point Process is a post-processing method for diversified recommendations. It produces a diverse set of items from the recommended items generated by DCN with greedy selection.
    \item \textbf{SSD} \cite{huang2021sliding_ssd}: It derives a time series analysis method called Sliding Spectrum Decomposition that better captures users’ perception of diversity in long sequences scenario.
    \item \textbf{MMR} \cite{carbonell1998use_mmr}: We implement the diversity scoring function following MMR which use a coarse-grained item-level diversity through re-ranking. 
\end{itemize}

\subsubsection{\textbf{Hyper-parameters and training details}.}
We implement the DCN model with three hidden layers MLP, and the activation function is ReLU. We optimize all models with Adam \cite{kingma2014adam} optimizer with batch sizes of $\{128,256\}$ on two datasets. We use grid search to find the optimal hyperparameters. In the CDM framework, $\lambda$ and $\gamma$, which balances accuracy and diversity, are searched in the range of $\{0.0, 0.1, ..., 0.25\}$. The learning rate is searched in $\{1e-3,1e-4,1e-5\}$. To avoid overfitting, we set the dropout \cite{srivastava2014dropout} to 0.25 and the patience of earlystop to 10 epochs. 

\subsection{Offline Comparison}

\begin{table*}[!th]
\caption{Overall Top-$K$ performance of different methods on JingDong and KuaiShou. Metric@$K$ denotes the corresponding Top-$K$ recommendation performance on this metric. RECALL and MRR focus on the accuracy of the recommended results, ILAD measures the diversity of the user-level recommendation list, and CC assesses the diversity at the system level. For each dataset, bold scores indicate the best in each column, underlined scores indicate the best baseline, and $^{+/-}$ represents whether the value is higher/lower compared to the DCN model. For all metrics, the higher the result, the better.}

\begin{tabular}{c|c|cccc|cccc}
\toprule
\multirow{7}{*}{JingDong} & Model & RECALL@3        & MRR@3           & ILAD@3          & CC@3            & RECALL@5        & MRR@5           & ILAD@5          & CC@5            \\ \cline{2-10} 
                          & DCN   & 0.0438\phantom{$^{-}$}           & 0.1001\phantom{$^{-}$}          & 0.2372\phantom{$^{-}$}          & 0.6943\phantom{$^{-}$}          & {\ul 0.0653}\phantom{$^{-}$}    & {\ul 0.1124}\phantom{$^{-}$}    & 0.2464\phantom{$^{-}$}          & 0.7210\phantom{$^{-}$}          \\
                          & IPW   & 0.0421$^{-}$    & 0.0993$^{-}$    & 0.2367$^{-}$    & 0.6939$^{-}$    & 0.0658$^{+}$    & 0.1138$^{+}$    & 0.2477$^{+}$    & 0.7311$^{+}$    \\
                          & DPP   & 0.0384$^{-}$    & 0.0929$^{-}$    & \textbf{0.2785}$^{+}$ & \textbf{0.7927}$^{+}$ & 0.0471$^{-}$    & 0.0830$^{-}$    & \textbf{0.3025}$^{+}$ & \textbf{0.8329}$^{+}$ \\
                          & SSD   & 0.0407$^{-}$    & 0.0955$^{-}$    & 0.2571$^{+}$    & { 0.7148}$^{+}$    & 0.0598$^{-}$    & 0.0902$^{-}$    & 0.2718$^{+}$    & {\ul 0.8016}$^{+}$    \\
                          & MMR   & {\ul 0.0459}$^{+}$ & {\ul 0.1038}$^{+}$ & 0.2493$^{+}$    & 0.7061$^{+}$    & 0.0637$^{-}$    & 0.1104$^{-}$    & 0.2609$^{-}$    & 0.7642$^{-}$    \\
                          & CDM   & \textbf{0.0488}$^{+}$    & \textbf{ 0.1074}$^{+}$    & {\ul 0.2611}$^{+}$    & {\ul 0.7342}$^{+}$    & \textbf{0.0713}$^{+}$ & \textbf{0.1192}$^{+}$ & {\ul 0.2785}$^{+}$    & 0.7874$^{+}$    \\ \midrule
\multirow{7}{*}{KuaiShou} & Model & RECALL@20       & MRR@20          & ILAD@20         & CC@20           & RECALL@50       & MRR@50          & ILAD@50         & CC@50           \\ \cline{2-10} 
                          & DCN   & 0.1057\phantom{$^{-}$}          & 0.0638\phantom{$^{-}$}          & 0.3922\phantom{$^{-}$}          & 0.7418\phantom{$^{-}$}          & 0.1931\phantom{$^{-}$}          & 0.0796\phantom{$^{-}$}          & 0.4502\phantom{$^{-}$}          & 0.8223\phantom{$^{-}$}          \\
                          & IPW   & 0.1006$^{-}$    & 0.0594$^{-}$    & 0.4019$^{+}$    & 0.7629$^{+}$    & 0.1828$^{-}$    & 0.0733$^{-}$    & 0.4719$^{+}$    & 0.8524$^{+}$    \\
                          & DPP   & 0.0842$^{-}$    & 0.0513$^{-}$    & \textbf{0.4917}$^{+}$ & \textbf{0.8942}$^{+}$ & 0.1629$^{-}$    & 0.0608$^{-}$    & \textbf{0.5719}$^{+}$ & \textbf{0.9640}$^{+}$ \\
                          & SSD   & 0.0891$^{-}$    & 0.0507$^{-}$    & {\ul 0.4555}$^{+}$    & 0.8007$^{+}$    & 0.1663$^{-}$    & 0.0656$^{-}$    & 0.5157$^{+}$    & 0.8931$^{+}$    \\
                          & MMR   & {\ul 0.1083}$^{+}$    & {\ul 0.0652}$^{+}$    & 0.4268$^{+}$    & 0.8248$^{+}$    & {\ul 0.2042}$^{+}$    & {\ul 0.0805}$^{+}$    & 0.5294$^{+}$    & 0.8902$^{+}$    \\
                          & CDM   & \textbf{0.1205}$^{+}$ & \textbf{0.0720}$^{+}$ & 0.4426$^{+}$   & {\ul 0.8461}$^{+}$    & \textbf{0.2219}$^{+}$ & \textbf{0.0894}$^{+}$ & {\ul 0.5503}$^{+}$    & {\ul 0.9147}$^{+}$    \\ \bottomrule
\end{tabular}
\end{table*}

In this section, we first evaluate all algorithms based on the accuracy and diversity as two key metrics. Table 2 presents the results of our offline experiments on two industrial dataset. Based on these results, we have the following findings:
(1) Compared to the base DCN model, all algorithms can enhance both user-level and system-level diversity. (2) IPW re-weights the loss based on category distribution to improve diversity. However, it dose not consistently outperform DCN in recommendation diversity or accuracy, due to the inaccurate estimation and high variance of propensity scores \cite{yao2021survey_ipw}. (3) 
DPP and SSD exhibit relatively similar performance. While they achieve higher diversity metrics, their accuracy drops more significantly compared to DCN. Some previous works \cite{zhang2023disentangled,lin2022feature} suggest that both of them assume that more orthogonality of recommended items would result in larger diversity in recommendation results. However, stronger orthogonality tends to lead to a significant decrease in accuracy, making it hard for them to effectively balance accuracy and diversity. (4) Consistent with previous findings \cite{zhang2023disentangled,lin2022feature}, MMR has the best performance of all baselines. We believe that under personalized recommendation, it is more appropriate to use embedded dot product to compute diversity scores than orthogonality. (5) Our proposed CDM effectively improved both recommendation accuracy and diversity on two datasets compared to the base DCN model. The effectiveness of CDM can be attributed to three following factors:
\begin{itemize}
    \item \textbf{Generalization}: CDM extracts diversity scores for each user-item pair as supervised signals from MMR, replacing the iterative strategy of MMR with an end-to-end manner, which may have better generalization.
    \item \textbf{Capacity}: However, just relying on such signals often falls short of surpassing teacher model. More significantly, CDM enhances its performance by modeling the distribution of candidate items produced by \textit{pre-rank} model. This enables each item to be aware of  relevant/competing items in the candidate set, thereby increasing the recommendation model's capacity \cite{jd_context}.
    \item \textbf{Information Exchange}: Additionally, the shared bottom embeddings benefit from additional update path, facilitating information exchange between the recommendation model and branch model. This, to some extent, can enhance the accuracy of the original recommendation model.
\end{itemize}

\subsection{Further Study}
In this section, we conduct more detailed analysis experiments to demonstrate the effectiveness of CDM. 
\subsubsection{\textbf{The impact of context modeling}} To simultaneously improve recommendation accuracy and diversity, CDM introduces an additional branch model on top of the recommendation backbone model. Next, we investigate the impact of various components of CDM on the performance. Specifically, we examine the following variables of the framework and compare them through offline experiments, which  showed in Table \ref{tab:further}.

\begin{table}[]
\caption{Performance comparison of different variants.}
\begin{tabular}{c|c|cccc}
\toprule
Dataset                                                                        & Module   & Recall          & MMR             & ILAD            & CC              \\ \hline
\multirow{3}{*}{\begin{tabular}[c]{@{}c@{}}JingDong\\ Metrics@5\end{tabular}}  & CDM      & \textbf{0.0713} & \textbf{0.1192} & 0.2785    &  \textbf{0.7874}    \\
                                                                               & -InfoNCE & 0.0677          & 0.1139          & 0.2693          & 0.7785          \\
                                                                               & -CCE     & 0.0624          & 0.1110          & \textbf{0.2792} & 0.7862 \\ \midrule
\multirow{3}{*}{\begin{tabular}[c]{@{}c@{}}KuaiShou\\ Metrics@50\end{tabular}} & CDM      & \textbf{0.2219} & \textbf{0.0894} & \textbf{0.5503}    &  \textbf{0.9147}    \\
                                                                               & -InfoNCE & 0.2095          & 0.0803          & 0.5257          & 0.8920          \\
                                                                               & -CCE     & 0.2036          & 0.0786          & 0.5223 & 0.8884 \\ \bottomrule
\end{tabular}\label{tab:further}
\end{table}

In CDM, the Contrastive Context Encoder (CCE) categorizes candidate items into positive and negative classes and then learns positive and negative contexts through a contrastive learning approach. From Table 2, it can be observed that when the InfoNCE loss, which controls context learning, is removed, CDM experiences a decrease in both accuracy and diversity. One possible explanation is that without this constraint, the context may capture redundant information, leading to reduced distinguishability between positive and negative contexts.

To further investigate the CCE module, we chose to directly model the input sequence using the target attention, thus eliminating the need to distinguish context. It was found that CDM's performance significantly deteriorated across all metrics. This suggests that CCE's strategy of contrastive context modeling enables each item to perceive information about its similar or competing items when being scored. This approach allows point-wise training to incorporate global candidate information, making it simpler and more effective.

\subsubsection{\textbf{The impact of trade-off parameter $\gamma$}}

\begin{figure}[t]
	\centering
	\subfigure[Top@5 on JingDong.]{\includegraphics[height=.38\linewidth]{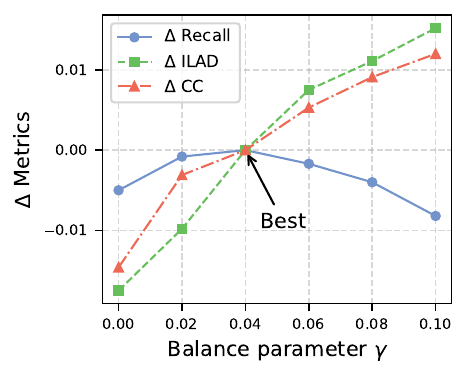}}\hspace{5pt}
	\subfigure[Top@50 on KuaiShou.]{\includegraphics[height=.38\linewidth]{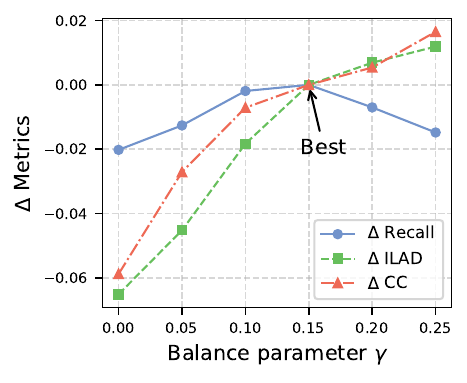}}
	\caption{Performance on accuracy and diversity when  varying $\gamma$.}
	\label{fig:tuning}
\end{figure}

In our framework, each item has two scores: accuracy and diversity. CDM combines two scores linearly, using $\text{ACC}(u,i)+ \gamma*\text{DIV}(u,i)$, where $\gamma$ is the hyper-parameter to balance accuracy and diversity. Fig \ref{fig:tuning} illustrates performance tuning on two datasets. The y-axis represents the difference between the metric value at that $\gamma$ and its best value, mathematically formulated as:$\Delta \text{Metric}(\gamma) = \text{Metric}(\gamma)-\text{Metric}(\gamma_{best})$. As $\gamma$ ascends, RECALL demonstrates a parabolic tendency, first elevating to a peak and then diminishing, signaling an optimal $\gamma$ value at which RECALL is maximized, which may due to the ability of CDM to perceive relevant/competing information. In contrast, ILAD and CC consistently exhibits an upward trajectory with the rise of $\gamma$, reflecting a proportional relation that highlights the enhanced emphasis on diversity with a larger $\gamma$. The results indicate that it is crucial to introduce a proper amount of diversity into the Top-$K$ list to improve the joint utility of accuracy and diversity for feed recommendation. Therefore, it is crucial to adjust the parameter $\gamma$ to achieve the optimal balance of accuracy and diversity to provide the best experience for users in practical applications.

\subsection{Online \textit{A/B} Test}

To validate the efficacy of CDM, we deploy it within the feed recommendation pages of two applications of KuaiShou for online \textit{A/B} test. The branch network is integrated into the online two-tower recommendation model, which additionally receives the output (i.e. candidate items) from the \textit{pre-rank} model, ensuring consistency with the offline experiments.

We evaluate the performance of CDM within a practical application context of \textit{Kuaishou}, referencing primarily the subsequent three metrics: (1) \textbf{Watch Time}: a crucial measure reflecting user engagement and satisfaction. (2) \textbf{\#Vertical Categories}: it signifies the diversity of presented content, representing the number of unique video categories showed to all users. (3) \textbf{Clustering Coefficient}: this measures the variability in user interaction with the recommended content, quantifying the difference of the number of videos to the number of unique video categories per session on average. A lower value suggests a concentration of user interactions within more categories, illustrating increased diversity in content consumption.


The \textit{A/B} test spanned ten consecutive days, with the average performance on the three aforementioned metrics tabulated in Table 4, allowing us to formulate the ensuing conclusions. Initially, CDM manifests a marked enhancement in Watch Time, corroborating the model’s capability to bolster user loyalty and engagement. Subsequently, the augmentation in the quantity of Vertical Categories coupled with the diminution in the Clustering Coefficient of observed videos attests to the promotion of diversity within the recommended outcomes. In essence, CDM fosters both diversity in recommendations and user engagement by strategically incorporating pipeline information, enabling models in the \textit{rank} stage to assimilate preferences from upstream models and data regarding candidates. This substantiates that our framework orchestrates an optimal balance between precision and diversity.

\begin{table}[t]
\caption{Online A/B test on two KuaiShou APP pages. We show the relative performance of CDM. The grey square brackets represent the 95\% confidence intervals for online metrics.}
\scalebox{0.9}{
\begin{tabular}{c|ccc}
\toprule
  Pages        & Watch Time & \#Vertical Categories & Clustering Coefficient \\ \hline
Main      & +0.406\%   & +0.188\%            & -0.957\%          \\ 
 & \textcolor{gray}{[0.33\%, 0.48\%]}   & \textcolor{gray}{[0.14\%, 0.23\%]}          & \textcolor{gray}{[-1.00\%, -0.91\%]}       \\
\hline
Lite & +0.412\%   & +0.264\%            & -1.106\%          \\ 
 & \textcolor{gray}{[0.17\%, 0.65\%]}   & \textcolor{gray}{[0.11\%, 0.45\%]}          & \textcolor{gray}{[-1.25\%, -0.87\%]}       \\
\bottomrule
\end{tabular}
}
\end{table}

\section{Related Work}
The discussion on diversified recommendation has been extensive within the research community, highlighting the recognition of factors beyond recommendation accuracy as pivotal in elevating overall user satisfaction \cite{wu2019_div_survey,kunaver2017diversity,hezz1,hezz2,xiao2023know,xiao2023user,xiao2024SmartGuard}.
This awareness has led to the emergence of the accuracy-diversity dilemma, a scenario where the attainment of higher accuracy often compromises diversity, and vice versa \cite{wang2021deconfounded,zheng2021dgcn}.
A primary cause of this dilemma has been the historical focus on accuracy \cite{wang2017deep_dcn, zhang2023disentangled}.
To improve user satisfaction, two distinct diversity problems have been delineated: aggregation diversity \cite{ge2010beyond,zhang2021model} and individual diversity \cite{cheng2017learning,wilhelm2018practical,sha2016framework,abdool2020managing,ashkan2015optimal,xu2023multi,chen2018fast_dpp}.
Aggregation diversity strives to diversify recommendations across the entire user, thereby enhancing the coverage of the recommender system.
On the other hand, individual diversity focuses on diversifying recommendations for a specific user, endeavoring to harmonize accuracy and diversity by ensuring a range of different items within a single user's recommendation list \cite{wu2019_div_survey}.

In this paper, we focus on the issue of individual diversity. A classical approach is the greedy-based iterative selection algorithms \cite{wu2019_div_survey}.
For instance, Maximal Marginal Relevance (MMR) \cite{carbonell1998use_mmr} iteratively selects items, taking into account both accuracy and their similarity to previously selected items.
Determinantal point process (DPP) \cite{gillenwater2014expectation} leverages a symmetric matrix to represent item qualities and pairwise similarities.
Sliding Spectrum Decomposition (SSD) \cite{huang2021sliding_ssd} proposes a time series analysis technique to include out-of-window items into the measurement of diversity to increase the diversity of a long recommendation sequence and alleviate the long tail effect as well.
Feature Disentanglement Self-Balancing (FDSB) aims to improve the diversity of \textit{re-ranking} stage by refining MMR in relevant recommendation. 
\cite{xu2023multi} proposes a general re-ranking framework named Multi-factor Sequential Re-ranking with Perception-Aware Diversification (MPAD) to jointly optimize accuracy and diversity for feed recommendation with DPP. 
Furthermore, some approaches take into account the temporal dimension of recommendations and utilize time series analysis techniques to model the diversity of recommendation sequences \cite{kaptein2009result, wilhelm2018practical}. These methods endeavor to capture information pertaining to out-of-window items, thereby aligning more closely with users' perception.
Different from the above approaches, our work is the first to consider both diversity and accuracy in \textit{rank} stage for recommendation with a flexible end-to-end algorithm. 

\section{Conclusion}

In conclusion, this paper highlights the critical role of diversity, on par with accuracy, in enhancing user experience in recommendation systems. Addressing the limitations of existing methods like DPP and MMR, which are constrained by quadratic complexity to the \textit{re-ranking} stage, we introduced the Contextual Distillation Model (CDM). Designed for efficient diversification across all recommendation stages, CDM leverages contextual information from candidate items and employs a contrastive context encoder to model diverse contexts effectively. Through offline and online evaluations, CDM demonstrated significant improvements in both recommendation quality and diversity, also ensuring efficiency.

\section*{Acknowledgment}
This work is partially supported in part by the National Natural Science Foundation of China (No. U2341229); the Key R\&D Project of Jilin Province (No. 20240304200SF);  and the International Cooperation Project of Jilin Province (No. 20220402009GH).

\bibliographystyle{ACM-Reference-Format}
\bibliography{reference}


\begin{thebibliography}{50}


\ifx \showCODEN    \undefined \def \showCODEN     #1{\unskip}     \fi
\ifx \showDOI      \undefined \def \showDOI       #1{#1}\fi
\ifx \showISBNx    \undefined \def \showISBNx     #1{\unskip}     \fi
\ifx \showISBNxiii \undefined \def \showISBNxiii  #1{\unskip}     \fi
\ifx \showISSN     \undefined \def \showISSN      #1{\unskip}     \fi
\ifx \showLCCN     \undefined \def \showLCCN      #1{\unskip}     \fi
\ifx \shownote     \undefined \def \shownote      #1{#1}          \fi
\ifx \showarticletitle \undefined \def \showarticletitle #1{#1}   \fi
\ifx \showURL      \undefined \def \showURL       {\relax}        \fi
\providecommand\bibfield[2]{#2}
\providecommand\bibinfo[2]{#2}
\providecommand\natexlab[1]{#1}
\providecommand\showeprint[2][]{arXiv:#2}

\bibitem[\protect\citeauthoryear{Abdool, Haldar, Ramanathan, Sax, Zhang, Manaswala, Yang, Turnbull, Zhang, and Legrand}{Abdool et~al\mbox{.}}{2020}]%
        {abdool2020managing}
\bibfield{author}{\bibinfo{person}{Mustafa Abdool}, \bibinfo{person}{Malay Haldar}, \bibinfo{person}{Prashant Ramanathan}, \bibinfo{person}{Tyler Sax}, \bibinfo{person}{Lanbo Zhang}, \bibinfo{person}{Aamir Manaswala}, \bibinfo{person}{Lynn Yang}, \bibinfo{person}{Bradley Turnbull}, \bibinfo{person}{Qing Zhang}, {and} \bibinfo{person}{Thomas Legrand}.} \bibinfo{year}{2020}\natexlab{}.
\newblock \showarticletitle{Managing diversity in airbnb search}. In \bibinfo{booktitle}{\emph{Proceedings of the 26th ACM SIGKDD International Conference on Knowledge Discovery \& Data Mining}}. \bibinfo{pages}{2952--2960}.
\newblock


\bibitem[\protect\citeauthoryear{Ashkan, Kveton, Berkovsky, and Wen}{Ashkan et~al\mbox{.}}{2015}]%
        {ashkan2015optimal}
\bibfield{author}{\bibinfo{person}{Azin Ashkan}, \bibinfo{person}{Branislav Kveton}, \bibinfo{person}{Shlomo Berkovsky}, {and} \bibinfo{person}{Zheng Wen}.} \bibinfo{year}{2015}\natexlab{}.
\newblock \showarticletitle{Optimal Greedy Diversity for Recommendation.}. In \bibinfo{booktitle}{\emph{IJCAI}}, Vol.~\bibinfo{volume}{15}. \bibinfo{pages}{1742--1748}.
\newblock


\bibitem[\protect\citeauthoryear{Bai, Wu, Hou, Cai, He, Zhou, Hong, and Wang}{Bai et~al\mbox{.}}{2024}]%
        {hezz2}
\bibfield{author}{\bibinfo{person}{Haoyue Bai}, \bibinfo{person}{Le Wu}, \bibinfo{person}{Min Hou}, \bibinfo{person}{Miaomiao Cai}, \bibinfo{person}{Zhuangzhuang He}, \bibinfo{person}{Yuyang Zhou}, \bibinfo{person}{Richang Hong}, {and} \bibinfo{person}{Meng Wang}.} \bibinfo{year}{2024}\natexlab{}.
\newblock \showarticletitle{Multimodality Invariant Learning for Multimedia-Based New Item Recommendation}.
\newblock \bibinfo{journal}{\emph{arXiv preprint arXiv:2405.15783}} (\bibinfo{year}{2024}).
\newblock


\bibitem[\protect\citeauthoryear{Carbonell and Goldstein}{Carbonell and Goldstein}{1998}]%
        {carbonell1998use_mmr}
\bibfield{author}{\bibinfo{person}{Jaime Carbonell} {and} \bibinfo{person}{Jade Goldstein}.} \bibinfo{year}{1998}\natexlab{}.
\newblock \showarticletitle{The use of MMR, diversity-based reranking for reordering documents and producing summaries}. In \bibinfo{booktitle}{\emph{Proceedings of the 21st annual international ACM SIGIR conference on Research and development in information retrieval}}. \bibinfo{pages}{335--336}.
\newblock


\bibitem[\protect\citeauthoryear{Chen, Zhang, and Zhou}{Chen et~al\mbox{.}}{2018}]%
        {chen2018fast_dpp}
\bibfield{author}{\bibinfo{person}{Laming Chen}, \bibinfo{person}{Guoxin Zhang}, {and} \bibinfo{person}{Eric Zhou}.} \bibinfo{year}{2018}\natexlab{}.
\newblock \showarticletitle{Fast greedy map inference for determinantal point process to improve recommendation diversity}.
\newblock \bibinfo{journal}{\emph{Advances in Neural Information Processing Systems}}  \bibinfo{volume}{31} (\bibinfo{year}{2018}).
\newblock


\bibitem[\protect\citeauthoryear{Cheng, Wang, Ma, Sun, and Xiong}{Cheng et~al\mbox{.}}{2017}]%
        {cheng2017learning}
\bibfield{author}{\bibinfo{person}{Peizhe Cheng}, \bibinfo{person}{Shuaiqiang Wang}, \bibinfo{person}{Jun Ma}, \bibinfo{person}{Jiankai Sun}, {and} \bibinfo{person}{Hui Xiong}.} \bibinfo{year}{2017}\natexlab{}.
\newblock \showarticletitle{Learning to recommend accurate and diverse items}. In \bibinfo{booktitle}{\emph{Proceedings of the 26th international conference on World Wide Web}}. \bibinfo{pages}{183--192}.
\newblock


\bibitem[\protect\citeauthoryear{Duan, Yu, Yin, Zhang, Luo, and Zhang}{Duan et~al\mbox{.}}{2019}]%
        {duan2019contrastive}
\bibfield{author}{\bibinfo{person}{Xiangyu Duan}, \bibinfo{person}{Hongfei Yu}, \bibinfo{person}{Mingming Yin}, \bibinfo{person}{Min Zhang}, \bibinfo{person}{Weihua Luo}, {and} \bibinfo{person}{Yue Zhang}.} \bibinfo{year}{2019}\natexlab{}.
\newblock \showarticletitle{Contrastive Attention Mechanism for Abstractive Sentence Summarization}.
\newblock \bibinfo{journal}{\emph{arXiv preprint arXiv:1910.13114}} (\bibinfo{year}{2019}).
\newblock


\bibitem[\protect\citeauthoryear{Fu, Miao, Zhang, Kuang, and Wu}{Fu et~al\mbox{.}}{2023}]%
        {fu2}
\bibfield{author}{\bibinfo{person}{Kairui Fu}, \bibinfo{person}{Qiaowei Miao}, \bibinfo{person}{Shengyu Zhang}, \bibinfo{person}{Kun Kuang}, {and} \bibinfo{person}{Fei Wu}.} \bibinfo{year}{2023}\natexlab{}.
\newblock \showarticletitle{End-to-End Optimization of Quantization-Based Structure Learning and Interventional Next-Item Recommendation}. In \bibinfo{booktitle}{\emph{CAAI International Conference on Artificial Intelligence}}. \bibinfo{pages}{415--429}.
\newblock


\bibitem[\protect\citeauthoryear{Fu, Zhang, Lv, Chen, and Li}{Fu et~al\mbox{.}}{2024}]%
        {fu1}
\bibfield{author}{\bibinfo{person}{Kairui Fu}, \bibinfo{person}{Shengyu Zhang}, \bibinfo{person}{Zheqi Lv}, \bibinfo{person}{Jingyuan Chen}, {and} \bibinfo{person}{Jiwei Li}.} \bibinfo{year}{2024}\natexlab{}.
\newblock \bibinfo{title}{DIET: Customized Slimming for Incompatible Networks in Sequential Recommendation}.
\newblock
\newblock
\showeprint{arXiv:2406.08804}


\bibitem[\protect\citeauthoryear{Ge, Delgado-Battenfeld, and Jannach}{Ge et~al\mbox{.}}{2010}]%
        {ge2010beyond}
\bibfield{author}{\bibinfo{person}{Mouzhi Ge}, \bibinfo{person}{Carla Delgado-Battenfeld}, {and} \bibinfo{person}{Dietmar Jannach}.} \bibinfo{year}{2010}\natexlab{}.
\newblock \showarticletitle{Beyond accuracy: evaluating recommender systems by coverage and serendipity}. In \bibinfo{booktitle}{\emph{Proceedings of the fourth ACM conference on Recommender systems}}. \bibinfo{pages}{257--260}.
\newblock


\bibitem[\protect\citeauthoryear{Gillenwater, Kulesza, Fox, and Taskar}{Gillenwater et~al\mbox{.}}{2014}]%
        {gillenwater2014expectation}
\bibfield{author}{\bibinfo{person}{Jennifer~A Gillenwater}, \bibinfo{person}{Alex Kulesza}, \bibinfo{person}{Emily Fox}, {and} \bibinfo{person}{Ben Taskar}.} \bibinfo{year}{2014}\natexlab{}.
\newblock \showarticletitle{Expectation-maximization for learning determinantal point processes}.
\newblock \bibinfo{journal}{\emph{Advances in Neural Information Processing Systems}}  \bibinfo{volume}{27} (\bibinfo{year}{2014}).
\newblock


\bibitem[\protect\citeauthoryear{He, Wang, Yang, Sun, Wu, Bai, Gong, Hong, and Zhang}{He et~al\mbox{.}}{2024}]%
        {hezz1}
\bibfield{author}{\bibinfo{person}{Zhuangzhuang He}, \bibinfo{person}{Yifan Wang}, \bibinfo{person}{Yonghui Yang}, \bibinfo{person}{Peijie Sun}, \bibinfo{person}{Le Wu}, \bibinfo{person}{Haoyue Bai}, \bibinfo{person}{Jinqi Gong}, \bibinfo{person}{Richang Hong}, {and} \bibinfo{person}{Min Zhang}.} \bibinfo{year}{2024}\natexlab{}.
\newblock \showarticletitle{Double Correction Framework for Denoising Recommendation}.
\newblock \bibinfo{journal}{\emph{arXiv preprint arXiv:2405.11272}} (\bibinfo{year}{2024}).
\newblock


\bibitem[\protect\citeauthoryear{Hinton, Vinyals, and Dean}{Hinton et~al\mbox{.}}{2015}]%
        {hinton2015distilling}
\bibfield{author}{\bibinfo{person}{Geoffrey Hinton}, \bibinfo{person}{Oriol Vinyals}, {and} \bibinfo{person}{Jeff Dean}.} \bibinfo{year}{2015}\natexlab{}.
\newblock \showarticletitle{Distilling the Knowledge in a Neural Network}.
\newblock \bibinfo{journal}{\emph{arXiv preprint arXiv:1503.02531}} (\bibinfo{year}{2015}).
\newblock
\urldef\tempurl%
\url{https://arxiv.org/abs/1503.02531}
\showURL{%
\tempurl}


\bibitem[\protect\citeauthoryear{Huang, Wang, Zhang, and Xu}{Huang et~al\mbox{.}}{2021}]%
        {huang2021sliding_ssd}
\bibfield{author}{\bibinfo{person}{Yanhua Huang}, \bibinfo{person}{Weikun Wang}, \bibinfo{person}{Lei Zhang}, {and} \bibinfo{person}{Ruiwen Xu}.} \bibinfo{year}{2021}\natexlab{}.
\newblock \showarticletitle{Sliding spectrum decomposition for diversified recommendation}. In \bibinfo{booktitle}{\emph{Proceedings of the 27th ACM SIGKDD Conference on Knowledge Discovery \& Data Mining}}. \bibinfo{pages}{3041--3049}.
\newblock


\bibitem[\protect\citeauthoryear{Jang, Gu, and Poole}{Jang et~al\mbox{.}}{2016}]%
        {jang2016categorical}
\bibfield{author}{\bibinfo{person}{Eric Jang}, \bibinfo{person}{Shixiang Gu}, {and} \bibinfo{person}{Ben Poole}.} \bibinfo{year}{2016}\natexlab{}.
\newblock \showarticletitle{Categorical Reparameterization with Gumbel-Softmax}.
\newblock \bibinfo{journal}{\emph{arXiv preprint arXiv:1611.01144}} (\bibinfo{year}{2016}).
\newblock


\bibitem[\protect\citeauthoryear{Kapoor, Kumar, Terveen, Konstan, and Schrater}{Kapoor et~al\mbox{.}}{2015}]%
        {kapoor2015like}
\bibfield{author}{\bibinfo{person}{Komal Kapoor}, \bibinfo{person}{Vikas Kumar}, \bibinfo{person}{Loren Terveen}, \bibinfo{person}{Joseph~A Konstan}, {and} \bibinfo{person}{Paul Schrater}.} \bibinfo{year}{2015}\natexlab{}.
\newblock \showarticletitle{" I like to explore sometimes" Adapting to Dynamic User Novelty Preferences}. In \bibinfo{booktitle}{\emph{Proceedings of the 9th ACM Conference on Recommender Systems}}. \bibinfo{pages}{19--26}.
\newblock


\bibitem[\protect\citeauthoryear{Kaptein, Koolen, and Kamps}{Kaptein et~al\mbox{.}}{2009}]%
        {kaptein2009result}
\bibfield{author}{\bibinfo{person}{R Kaptein}, \bibinfo{person}{M Koolen}, {and} \bibinfo{person}{J Kamps}.} \bibinfo{year}{2009}\natexlab{}.
\newblock \showarticletitle{Result Diversity and Entity Ranking Experiments: Anchors, Links, Text and Wikipedia, University of Amsterdam}.
\newblock \bibinfo{journal}{\emph{NIST Special Publication}} (\bibinfo{year}{2009}), \bibinfo{pages}{500--278}.
\newblock


\bibitem[\protect\citeauthoryear{Kingma and Ba}{Kingma and Ba}{2014}]%
        {kingma2014adam}
\bibfield{author}{\bibinfo{person}{Diederik~P Kingma} {and} \bibinfo{person}{Jimmy Ba}.} \bibinfo{year}{2014}\natexlab{}.
\newblock \showarticletitle{Adam: A Method for Stochastic Optimization}.
\newblock \bibinfo{journal}{\emph{arXiv preprint arXiv:1412.6980}} (\bibinfo{year}{2014}).
\newblock


\bibitem[\protect\citeauthoryear{Kunaver and Po{\v{z}}rl}{Kunaver and Po{\v{z}}rl}{2017}]%
        {kunaver2017diversity}
\bibfield{author}{\bibinfo{person}{Matev{\v{z}} Kunaver} {and} \bibinfo{person}{Toma{\v{z}} Po{\v{z}}rl}.} \bibinfo{year}{2017}\natexlab{}.
\newblock \showarticletitle{Diversity in recommender systems--A survey}.
\newblock \bibinfo{journal}{\emph{Knowledge-based systems}}  \bibinfo{volume}{123} (\bibinfo{year}{2017}), \bibinfo{pages}{154--162}.
\newblock


\bibitem[\protect\citeauthoryear{Lin, Wang, Mao, Zhao, Wang, Jiang, and Wen}{Lin et~al\mbox{.}}{2022}]%
        {lin2022feature}
\bibfield{author}{\bibinfo{person}{Zihan Lin}, \bibinfo{person}{Hui Wang}, \bibinfo{person}{Jingshu Mao}, \bibinfo{person}{Wayne~Xin Zhao}, \bibinfo{person}{Cheng Wang}, \bibinfo{person}{Peng Jiang}, {and} \bibinfo{person}{Ji-Rong Wen}.} \bibinfo{year}{2022}\natexlab{}.
\newblock \showarticletitle{Feature-aware diversified re-ranking with disentangled representations for relevant recommendation}. In \bibinfo{booktitle}{\emph{Proceedings of the 28th ACM SIGKDD Conference on Knowledge Discovery and Data Mining}}. \bibinfo{pages}{3327--3335}.
\newblock


\bibitem[\protect\citeauthoryear{Minoux}{Minoux}{2005}]%
        {minoux2005accelerated}
\bibfield{author}{\bibinfo{person}{Michel Minoux}.} \bibinfo{year}{2005}\natexlab{}.
\newblock \showarticletitle{Accelerated greedy algorithms for maximizing submodular set functions}. In \bibinfo{booktitle}{\emph{Optimization Techniques: Proceedings of the 8th IFIP Conference on Optimization Techniques W{\"u}rzburg, September 5--9, 1977}}. Springer, \bibinfo{pages}{234--243}.
\newblock


\bibitem[\protect\citeauthoryear{Oord, Li, and Vinyals}{Oord et~al\mbox{.}}{2018}]%
        {oord2018representation}
\bibfield{author}{\bibinfo{person}{Aaron van~den Oord}, \bibinfo{person}{Yazhe Li}, {and} \bibinfo{person}{Oriol Vinyals}.} \bibinfo{year}{2018}\natexlab{}.
\newblock \showarticletitle{Representation learning with contrastive predictive coding}.
\newblock \bibinfo{journal}{\emph{arXiv preprint arXiv:1807.03748}} (\bibinfo{year}{2018}).
\newblock


\bibitem[\protect\citeauthoryear{Pl{\"o}tz and Roth}{Pl{\"o}tz and Roth}{2018}]%
        {plotz2018neural}
\bibfield{author}{\bibinfo{person}{Tobias Pl{\"o}tz} {and} \bibinfo{person}{Stefan Roth}.} \bibinfo{year}{2018}\natexlab{}.
\newblock \showarticletitle{Neural nearest neighbors networks}.
\newblock \bibinfo{journal}{\emph{Advances in Neural information processing systems}}  \bibinfo{volume}{31} (\bibinfo{year}{2018}).
\newblock


\bibitem[\protect\citeauthoryear{Saito, Yaginuma, Nishino, Sakata, and Nakata}{Saito et~al\mbox{.}}{2020}]%
        {saito2020unbiased_ipw}
\bibfield{author}{\bibinfo{person}{Yuta Saito}, \bibinfo{person}{Suguru Yaginuma}, \bibinfo{person}{Yuta Nishino}, \bibinfo{person}{Hayato Sakata}, {and} \bibinfo{person}{Kazuhide Nakata}.} \bibinfo{year}{2020}\natexlab{}.
\newblock \showarticletitle{Unbiased recommender learning from missing-not-at-random implicit feedback}. In \bibinfo{booktitle}{\emph{Proceedings of the 13th International Conference on Web Search and Data Mining}}. \bibinfo{pages}{501--509}.
\newblock


\bibitem[\protect\citeauthoryear{Sha, Wu, and Niu}{Sha et~al\mbox{.}}{2016}]%
        {sha2016framework}
\bibfield{author}{\bibinfo{person}{Chaofeng Sha}, \bibinfo{person}{Xiaowei Wu}, {and} \bibinfo{person}{Junyu Niu}.} \bibinfo{year}{2016}\natexlab{}.
\newblock \showarticletitle{A framework for recommending relevant and diverse items.}. In \bibinfo{booktitle}{\emph{IJCAI}}, Vol.~\bibinfo{volume}{16}. \bibinfo{pages}{3868--3874}.
\newblock


\bibitem[\protect\citeauthoryear{Song, Huang, Ouyang, and Wang}{Song et~al\mbox{.}}{2018}]%
        {song2018mask}
\bibfield{author}{\bibinfo{person}{Chunfeng Song}, \bibinfo{person}{Yan Huang}, \bibinfo{person}{Wanli Ouyang}, {and} \bibinfo{person}{Liang Wang}.} \bibinfo{year}{2018}\natexlab{}.
\newblock \showarticletitle{Mask-guided contrastive attention model for person re-identification}. In \bibinfo{booktitle}{\emph{Proceedings of the IEEE conference on computer vision and pattern recognition}}. \bibinfo{pages}{1179--1188}.
\newblock


\bibitem[\protect\citeauthoryear{Srivastava, Hinton, Krizhevsky, Sutskever, and Salakhutdinov}{Srivastava et~al\mbox{.}}{2014}]%
        {srivastava2014dropout}
\bibfield{author}{\bibinfo{person}{Nitish Srivastava}, \bibinfo{person}{Geoffrey Hinton}, \bibinfo{person}{Alex Krizhevsky}, \bibinfo{person}{Ilya Sutskever}, {and} \bibinfo{person}{Ruslan Salakhutdinov}.} \bibinfo{year}{2014}\natexlab{}.
\newblock \showarticletitle{Dropout: A simple way to prevent neural networks from overfitting}.
\newblock \bibinfo{journal}{\emph{The Journal of Machine Learning Research}} \bibinfo{volume}{15}, \bibinfo{number}{1} (\bibinfo{year}{2014}), \bibinfo{pages}{1929--1958}.
\newblock


\bibitem[\protect\citeauthoryear{Sun, Sun, Sha, Zhang, and Ong}{Sun et~al\mbox{.}}{2023}]%
        {sun1}
\bibfield{author}{\bibinfo{person}{Youchen Sun}, \bibinfo{person}{Zhu Sun}, \bibinfo{person}{Xiao Sha}, \bibinfo{person}{Jie Zhang}, {and} \bibinfo{person}{Yew~Soon Ong}.} \bibinfo{year}{2023}\natexlab{}.
\newblock \showarticletitle{Disentangling Motives behind Item Consumption and Social Connection for Mutually-enhanced Joint Prediction}. In \bibinfo{booktitle}{\emph{Proceedings of the 17th ACM Conference on Recommender Systems}}. \bibinfo{pages}{613--624}.
\newblock


\bibitem[\protect\citeauthoryear{Tang, Li, Ma, Gao, Wang, Jiang, Ma, Zhang, and Chen}{Tang et~al\mbox{.}}{2022}]%
        {shisong2022}
\bibfield{author}{\bibinfo{person}{Shisong Tang}, \bibinfo{person}{Qing Li}, \bibinfo{person}{Xiaoteng Ma}, \bibinfo{person}{Ci Gao}, \bibinfo{person}{Dingmin Wang}, \bibinfo{person}{Yong Jiang}, \bibinfo{person}{Qian Ma}, \bibinfo{person}{Aoyang Zhang}, {and} \bibinfo{person}{Hechang Chen}.} \bibinfo{year}{2022}\natexlab{}.
\newblock \showarticletitle{Knowledge-based temporal fusion network for interpretable online video popularity prediction}. In \bibinfo{booktitle}{\emph{Proceedings of the ACM Web Conference 2022}}. \bibinfo{pages}{2879--2887}.
\newblock


\bibitem[\protect\citeauthoryear{Tang, Li, Wang, Gao, Xiao, Zhao, Jiang, Ma, and Zhang}{Tang et~al\mbox{.}}{2023}]%
        {shisong2023}
\bibfield{author}{\bibinfo{person}{Shisong Tang}, \bibinfo{person}{Qing Li}, \bibinfo{person}{Dingmin Wang}, \bibinfo{person}{Ci Gao}, \bibinfo{person}{Wentao Xiao}, \bibinfo{person}{Dan Zhao}, \bibinfo{person}{Yong Jiang}, \bibinfo{person}{Qian Ma}, {and} \bibinfo{person}{Aoyang Zhang}.} \bibinfo{year}{2023}\natexlab{}.
\newblock \showarticletitle{Counterfactual Video Recommendation for Duration Debiasing}. In \bibinfo{booktitle}{\emph{Proceedings of the 29th ACM SIGKDD Conference on Knowledge Discovery and Data Mining}}. \bibinfo{pages}{4894--4903}.
\newblock


\bibitem[\protect\citeauthoryear{Toth}{Toth}{2000}]%
        {toth2000optimization}
\bibfield{author}{\bibinfo{person}{Paolo Toth}.} \bibinfo{year}{2000}\natexlab{}.
\newblock \showarticletitle{Optimization engineering techniques for the exact solution of NP-hard combinatorial optimization problems}.
\newblock \bibinfo{journal}{\emph{European journal of operational research}} \bibinfo{volume}{125}, \bibinfo{number}{2} (\bibinfo{year}{2000}), \bibinfo{pages}{222--238}.
\newblock


\bibitem[\protect\citeauthoryear{Van~Erven and Harremos}{Van~Erven and Harremos}{2014}]%
        {van2014renyi}
\bibfield{author}{\bibinfo{person}{Tim Van~Erven} {and} \bibinfo{person}{Peter Harremos}.} \bibinfo{year}{2014}\natexlab{}.
\newblock \showarticletitle{R{\'e}nyi divergence and Kullback-Leibler divergence}.
\newblock \bibinfo{journal}{\emph{IEEE Transactions on Information Theory}} \bibinfo{volume}{60}, \bibinfo{number}{7} (\bibinfo{year}{2014}), \bibinfo{pages}{3797--3820}.
\newblock


\bibitem[\protect\citeauthoryear{Vaswani, Shazeer, Parmar, Uszkoreit, Jones, Gomez, Kaiser, and Polosukhin}{Vaswani et~al\mbox{.}}{2017}]%
        {vaswani2017attention}
\bibfield{author}{\bibinfo{person}{Ashish Vaswani}, \bibinfo{person}{Noam Shazeer}, \bibinfo{person}{Niki Parmar}, \bibinfo{person}{Jakob Uszkoreit}, \bibinfo{person}{Llion Jones}, \bibinfo{person}{Aidan~N Gomez}, \bibinfo{person}{{\L}ukasz Kaiser}, {and} \bibinfo{person}{Illia Polosukhin}.} \bibinfo{year}{2017}\natexlab{}.
\newblock \showarticletitle{Attention is all you need}.
\newblock \bibinfo{journal}{\emph{Advances in neural information processing systems}}  \bibinfo{volume}{30} (\bibinfo{year}{2017}).
\newblock


\bibitem[\protect\citeauthoryear{Wang, Fu, Fu, and Wang}{Wang et~al\mbox{.}}{2017}]%
        {wang2017deep_dcn}
\bibfield{author}{\bibinfo{person}{Ruoxi Wang}, \bibinfo{person}{Bin Fu}, \bibinfo{person}{Gang Fu}, {and} \bibinfo{person}{Mingliang Wang}.} \bibinfo{year}{2017}\natexlab{}.
\newblock \showarticletitle{Deep \& cross network for ad click predictions}.
\newblock In \bibinfo{booktitle}{\emph{Proceedings of the ADKDD'17}}. \bibinfo{pages}{1--7}.
\newblock


\bibitem[\protect\citeauthoryear{Wang, Feng, He, Wang, and Chua}{Wang et~al\mbox{.}}{2021}]%
        {wang2021deconfounded}
\bibfield{author}{\bibinfo{person}{Wenjie Wang}, \bibinfo{person}{Fuli Feng}, \bibinfo{person}{Xiangnan He}, \bibinfo{person}{Xiang Wang}, {and} \bibinfo{person}{Tat-Seng Chua}.} \bibinfo{year}{2021}\natexlab{}.
\newblock \showarticletitle{Deconfounded recommendation for alleviating bias amplification}. In \bibinfo{booktitle}{\emph{Proceedings of the 27th ACM SIGKDD Conference on Knowledge Discovery \& Data Mining}}. \bibinfo{pages}{1717--1725}.
\newblock


\bibitem[\protect\citeauthoryear{Wilhelm, Ramanathan, Bonomo, Jain, Chi, and Gillenwater}{Wilhelm et~al\mbox{.}}{2018}]%
        {wilhelm2018practical}
\bibfield{author}{\bibinfo{person}{Mark Wilhelm}, \bibinfo{person}{Ajith Ramanathan}, \bibinfo{person}{Alexander Bonomo}, \bibinfo{person}{Sagar Jain}, \bibinfo{person}{Ed~H Chi}, {and} \bibinfo{person}{Jennifer Gillenwater}.} \bibinfo{year}{2018}\natexlab{}.
\newblock \showarticletitle{Practical diversified recommendations on youtube with determinantal point processes}. In \bibinfo{booktitle}{\emph{Proceedings of the 27th ACM International Conference on Information and Knowledge Management}}. \bibinfo{pages}{2165--2173}.
\newblock


\bibitem[\protect\citeauthoryear{Wu, Liu, Miao, Zhao, Guan, and Tang}{Wu et~al\mbox{.}}{2019a}]%
        {wu2019_div_survey}
\bibfield{author}{\bibinfo{person}{Qiong Wu}, \bibinfo{person}{Yong Liu}, \bibinfo{person}{Chunyan Miao}, \bibinfo{person}{Yin Zhao}, \bibinfo{person}{Lu Guan}, {and} \bibinfo{person}{Haihong Tang}.} \bibinfo{year}{2019}\natexlab{a}.
\newblock \showarticletitle{Recent advances in diversified recommendation}.
\newblock \bibinfo{journal}{\emph{arXiv preprint arXiv:1905.06589}} (\bibinfo{year}{2019}).
\newblock


\bibitem[\protect\citeauthoryear{Wu, Wang, Li, and Wang}{Wu et~al\mbox{.}}{2019b}]%
        {wu2019dynamic}
\bibfield{author}{\bibinfo{person}{Qingyun Wu}, \bibinfo{person}{Huazheng Wang}, \bibinfo{person}{Yanen Li}, {and} \bibinfo{person}{Hongning Wang}.} \bibinfo{year}{2019}\natexlab{b}.
\newblock \showarticletitle{Dynamic ensemble of contextual bandits to satisfy users' changing interests}. In \bibinfo{booktitle}{\emph{The World Wide Web Conference}}. \bibinfo{pages}{2080--2090}.
\newblock


\bibitem[\protect\citeauthoryear{Xiao, Xu, Zou, Li, Zhao, Fang, Li, Tang, Li, Zuo, Hu, Jiang, Weng, and Lyv.R}{Xiao et~al\mbox{.}}{2024}]%
        {xiao2024SmartGuard}
\bibfield{author}{\bibinfo{person}{Jingyu Xiao}, \bibinfo{person}{Zhiyao Xu}, \bibinfo{person}{Qingsong Zou}, \bibinfo{person}{Qing Li}, \bibinfo{person}{Dan Zhao}, \bibinfo{person}{Dong Fang}, \bibinfo{person}{Ruoyu Li}, \bibinfo{person}{Wenxin Tang}, \bibinfo{person}{Kang Li}, \bibinfo{person}{Xudong Zuo}, \bibinfo{person}{Penghui Hu}, \bibinfo{person}{Yong Jiang}, \bibinfo{person}{Zixuan Weng}, {and} \bibinfo{person}{Michael Lyv.R}.} \bibinfo{year}{2024}\natexlab{}.
\newblock \showarticletitle{Make Your Home Safe: Time-aware Unsupervised User Behavior Anomaly Detection in Smart Homes via Loss-guided Mask}.
\newblock \bibinfo{journal}{\emph{arXiv preprint arXiv:2406.10928}} (\bibinfo{year}{2024}).
\newblock
\urldef\tempurl%
\url{https://arxiv.org/abs/2406.10928}
\showURL{%
\tempurl}


\bibitem[\protect\citeauthoryear{Xiao, Zou, Li, Zhao, Li, Tang, Zhou, and Jiang}{Xiao et~al\mbox{.}}{2023a}]%
        {xiao2023user}
\bibfield{author}{\bibinfo{person}{Jingyu Xiao}, \bibinfo{person}{Qingsong Zou}, \bibinfo{person}{Qing Li}, \bibinfo{person}{Dan Zhao}, \bibinfo{person}{Kang Li}, \bibinfo{person}{Wenxin Tang}, \bibinfo{person}{Runjie Zhou}, {and} \bibinfo{person}{Yong Jiang}.} \bibinfo{year}{2023}\natexlab{a}.
\newblock \showarticletitle{User Device Interaction Prediction via Relational Gated Graph Attention Network and Intent-aware Encoder}. In \bibinfo{booktitle}{\emph{Proceedings of the 2023 International Conference on Autonomous Agents and Multiagent Systems (AAMAS)}}. \bibinfo{pages}{1634--1642}.
\newblock


\bibitem[\protect\citeauthoryear{Xiao, Zou, Li, Zhao, Li, Weng, Li, and Jiang}{Xiao et~al\mbox{.}}{2023b}]%
        {xiao2023know}
\bibfield{author}{\bibinfo{person}{Jingyu Xiao}, \bibinfo{person}{Qingsong Zou}, \bibinfo{person}{Qing Li}, \bibinfo{person}{Dan Zhao}, \bibinfo{person}{Kang Li}, \bibinfo{person}{Zixuan Weng}, \bibinfo{person}{Ruoyu Li}, {and} \bibinfo{person}{Yong Jiang}.} \bibinfo{year}{2023}\natexlab{b}.
\newblock \showarticletitle{I Know Your Intent: Graph-enhanced Intent-aware User Device Interaction Prediction via Contrastive Learning}.
\newblock \bibinfo{journal}{\emph{Proceedings of the ACM on Interactive, Mobile, Wearable and Ubiquitous Technologies (IMUWT/UbiComp)}} \bibinfo{volume}{7}, \bibinfo{number}{3} (\bibinfo{year}{2023}), \bibinfo{pages}{1--28}.
\newblock


\bibitem[\protect\citeauthoryear{Xie and Ermon}{Xie and Ermon}{2019}]%
        {xie2019reparameterizable}
\bibfield{author}{\bibinfo{person}{Sang~Michael Xie} {and} \bibinfo{person}{Stefano Ermon}.} \bibinfo{year}{2019}\natexlab{}.
\newblock \showarticletitle{Reparameterizable subset sampling via continuous relaxations}.
\newblock \bibinfo{journal}{\emph{arXiv preprint arXiv:1901.10517}} (\bibinfo{year}{2019}).
\newblock


\bibitem[\protect\citeauthoryear{Xu, Zhang, Zhang, Zhu, Su, Liu, Wong, Liao, and Choy}{Xu et~al\mbox{.}}{2015}]%
        {xu2015low}
\bibfield{author}{\bibinfo{person}{Mei-Feng Xu}, \bibinfo{person}{Hong Zhang}, \bibinfo{person}{Su Zhang}, \bibinfo{person}{Hugh~L Zhu}, \bibinfo{person}{Hui-Min Su}, \bibinfo{person}{Jian Liu}, \bibinfo{person}{Kam~Sing Wong}, \bibinfo{person}{Liang-Sheng Liao}, {and} \bibinfo{person}{Wallace~CH Choy}.} \bibinfo{year}{2015}\natexlab{}.
\newblock \showarticletitle{A low temperature gradual annealing scheme for achieving high performance perovskite solar cells with no hysteresis}.
\newblock \bibinfo{journal}{\emph{Journal of Materials Chemistry A}} \bibinfo{volume}{3}, \bibinfo{number}{27} (\bibinfo{year}{2015}), \bibinfo{pages}{14424--14430}.
\newblock


\bibitem[\protect\citeauthoryear{Xu, Chen, Wang, Yin, Shen, Wang, Huang, Lai, Zhuang, Ge, et~al\mbox{.}}{Xu et~al\mbox{.}}{2023}]%
        {xu2023multi}
\bibfield{author}{\bibinfo{person}{Yue Xu}, \bibinfo{person}{Hao Chen}, \bibinfo{person}{Zefan Wang}, \bibinfo{person}{Jianwen Yin}, \bibinfo{person}{Qijie Shen}, \bibinfo{person}{Dimin Wang}, \bibinfo{person}{Feiran Huang}, \bibinfo{person}{Lixiang Lai}, \bibinfo{person}{Tao Zhuang}, \bibinfo{person}{Junfeng Ge}, {et~al\mbox{.}}} \bibinfo{year}{2023}\natexlab{}.
\newblock \showarticletitle{Multi-factor Sequential Re-ranking with Perception-Aware Diversification}.
\newblock \bibinfo{journal}{\emph{arXiv preprint arXiv:2305.12420}} (\bibinfo{year}{2023}).
\newblock


\bibitem[\protect\citeauthoryear{Yao, Chu, Li, Li, Gao, and Zhang}{Yao et~al\mbox{.}}{2021}]%
        {yao2021survey_ipw}
\bibfield{author}{\bibinfo{person}{Liuyi Yao}, \bibinfo{person}{Zhixuan Chu}, \bibinfo{person}{Sheng Li}, \bibinfo{person}{Yaliang Li}, \bibinfo{person}{Jing Gao}, {and} \bibinfo{person}{Aidong Zhang}.} \bibinfo{year}{2021}\natexlab{}.
\newblock \showarticletitle{A survey on causal inference}.
\newblock \bibinfo{journal}{\emph{ACM Transactions on Knowledge Discovery from Data (TKDD)}} \bibinfo{volume}{15}, \bibinfo{number}{5} (\bibinfo{year}{2021}), \bibinfo{pages}{1--46}.
\newblock


\bibitem[\protect\citeauthoryear{Yu}{Yu}{2020}]%
        {yu2020tagnn}
\bibfield{author}{\bibinfo{person}{Feng Yu}.} \bibinfo{year}{2020}\natexlab{}.
\newblock \showarticletitle{TAGNN: Target Attentive Graph Neural Networks for Session-based Recommendation}. In \bibinfo{booktitle}{\emph{International ACM SIGIR Conference on Research and Development in Information Retrieval}}. \bibinfo{address}{Xi’an, China}.
\newblock
\urldef\tempurl%
\url{https://ar5iv.org/abs/2005.02844}
\showURL{%
\tempurl}


\bibitem[\protect\citeauthoryear{Zhang, Wang, and Li}{Zhang et~al\mbox{.}}{2023}]%
        {zhang2023disentangled}
\bibfield{author}{\bibinfo{person}{Xiaoying Zhang}, \bibinfo{person}{Hongning Wang}, {and} \bibinfo{person}{Hang Li}.} \bibinfo{year}{2023}\natexlab{}.
\newblock \showarticletitle{Disentangled Representation for Diversified Recommendations}. In \bibinfo{booktitle}{\emph{Proceedings of the Sixteenth ACM International Conference on Web Search and Data Mining}}. \bibinfo{pages}{490--498}.
\newblock


\bibitem[\protect\citeauthoryear{Zhang, Cheng, Yao, Yi, Hong, and Chi}{Zhang et~al\mbox{.}}{2021}]%
        {zhang2021model}
\bibfield{author}{\bibinfo{person}{Yin Zhang}, \bibinfo{person}{Derek~Zhiyuan Cheng}, \bibinfo{person}{Tiansheng Yao}, \bibinfo{person}{Xinyang Yi}, \bibinfo{person}{Lichan Hong}, {and} \bibinfo{person}{Ed~H Chi}.} \bibinfo{year}{2021}\natexlab{}.
\newblock \showarticletitle{A model of two tales: Dual transfer learning framework for improved long-tail item recommendation}. In \bibinfo{booktitle}{\emph{Proceedings of the web conference 2021}}. \bibinfo{pages}{2220--2231}.
\newblock


\bibitem[\protect\citeauthoryear{Zheng, Wang, Li, Chen, Liu, Lu, Zhao, Peng, Lin, and Shao}{Zheng et~al\mbox{.}}{2022}]%
        {jd_context}
\bibfield{author}{\bibinfo{person}{Kaifu Zheng}, \bibinfo{person}{Lu Wang}, \bibinfo{person}{Yu Li}, \bibinfo{person}{Xusong Chen}, \bibinfo{person}{Hu Liu}, \bibinfo{person}{Jing Lu}, \bibinfo{person}{Xiwei Zhao}, \bibinfo{person}{Changping Peng}, \bibinfo{person}{Zhangang Lin}, {and} \bibinfo{person}{Jingping Shao}.} \bibinfo{year}{2022}\natexlab{}.
\newblock \showarticletitle{Implicit User Awareness Modeling via Candidate Items for CTR Prediction in Search Ads}. In \bibinfo{booktitle}{\emph{Proceedings of the ACM Web Conference 2022}}. \bibinfo{pages}{246--255}.
\newblock


\bibitem[\protect\citeauthoryear{Zheng, Gao, Chen, Jin, and Li}{Zheng et~al\mbox{.}}{2021}]%
        {zheng2021dgcn}
\bibfield{author}{\bibinfo{person}{Yu Zheng}, \bibinfo{person}{Chen Gao}, \bibinfo{person}{Liang Chen}, \bibinfo{person}{Depeng Jin}, {and} \bibinfo{person}{Yong Li}.} \bibinfo{year}{2021}\natexlab{}.
\newblock \showarticletitle{DGCN: Diversified recommendation with graph convolutional networks}. In \bibinfo{booktitle}{\emph{Proceedings of the Web Conference 2021}}. \bibinfo{pages}{401--412}.
\newblock


\end{thebibliography}

\nocite{*}
\end{document}